\documentclass[a4paper,12pt]{article}
\usepackage{amsfonts,amssymb,color,amsmath,axodraw}

\usepackage{cite}

\long\def\note#1{\setbox0=\hbox{\large #1}
\ifdim\wd0<0.85\textwidth
\begin{center}\framebox%[\wd0]
{\color{blue}\quad\large #1\quad}\end{center}
\else
\begin{center}\framebox{\parbox{0.9\textwidth}
{\color{blue}\large #1}}\end{center}
\fi
}

\title{\bf Discrete R-symmetries and F-term Supersymmetry Breaking}
\author{\bf Pritibhajan Byakti\\
Saha Institute of Nuclear Physics\\
1/AF Bidhan-Nagar, Kolkata 700064, India.}
\date{}

\def\Eqn#1{Eq.\ (\ref{#1})}
\def\Eqs#1#2{Eq.\ (\ref{#1}) and (\ref{#2})}
\def\3Eqs#1#2#3{Eq.\ (\ref{#1}), (\ref{#2}) and (\ref{#3})}

\def\mod{\mathop{\rm mod}}

\def\adots{\mathinner{\mkern1mu\raise1pt\vbox{\kern7pt\hbox{.}}\mkern2mu
\raise4pt\hbox{.}\mkern2mu\raise7pt\hbox{.}\mkern1mu}}

\long\def\note#1{\setbox0=\hbox{\large #1}
\ifdim\wd0<0.85\textwidth
\begin{center}\framebox%[\wd0]
{\color{blue}\quad\large #1\quad}\end{center}
\else
\begin{center}\framebox{\parbox{0.9\textwidth}
{\color{blue}\large #1}}\end{center}
\fi
}

\begin{document}

\maketitle

\begin{abstract}
 We have shown that in a large number of generic and
renormalizable Wess-Zumino models existence of a $Z_n$ R-symmetry is
sufficient to break supersymmetry spontaneously. This implies that
existence of a $Z_n$ R-symmetry is a necessary condition for supersymmetry
breaking  in generic and renormalizable Wess-Zumino models. 
\end{abstract}

\section{Introduction}
In discussions of   F-term $N=1$ supersymmetry (SUSY) breaking,
R-symmetry plays a crucial role. Importance of a U(1) R-symmetry in spontaneous
SUSY breaking was clearly addressed in Ref.~\cite{Nelson:1993nf}. It was
shown  that, ``a continuous R-symmetry is a necessary condition
for spontaneous supersymmetry breaking and a spontaneously broken R-symmetry is
a sufficient condition, in models where the gauge dynamics can be integrated out
and in which the effective superpotential is a generic function consistent with
the symmetries of the theory.''

However, in this paper, we are going to show that there exist a large number of
generic and renormalizable Wess-Zumino(WZ) models where existence of a $Z_n$
R-symmetry is sufficient to break SUSY spontaneously. It is well
known  that if there is no R-symmetry, continuous or discrete, in WZ models
with canonical K\"{a}hler potentials, then the global  minima  preserve
SUSY~\cite{Nelson:1993nf,Argyres:susy2001.pdf}. With the help of the above two
results we can conclude that existence of a $Z_n$ R-symmetry is a necessary
condition for  breaking in a global minimum for generic and
renormalizable WZ models.

Models of F-term SUSY breaking with a $Z_n$ R-symmetry can be obtained  most
simply from the models of F-term SUSY breaking with U(1) R-symmetry by adding a
completely different and decoupled sector to break U(1) R-symmetry explicitly
down to a $Z_n$ R-symmetry. In sec~\ref{s:deformOR},  we illustrate this idea by
adding some more terms to the famous O'Raifeartaigh~\cite{O'Raifeartaigh:1975pr}
(O'R) model. However these models are trivial. In sec.~\ref{s:nontrivial}, we
explicitly discuss a non-trivial model where the superpotential cannot
be broken into such two non-interacting parts. We have also shown that this
model has a vacuum where SUSY as well as discrete R-symmetry is spontaneously
broken for large regions of parameter space. We then give some variations of
this model by adding more fields. In sec.~\ref{s:serises}, we identify the form
of these models, and then, three series of models are given where each series
contains a large  number of such models.

A note about notations.  For U(1) R-symmetry we use $R(\phi)$ to denote the U(1)
R-charge of any chiral scalar superfield $\phi$ in the normalization where the
R-charge of $\theta$ is 1.  We use $R_d(\theta)$ to denote $Z_n$ R-charge of
the superspace co-ordinate $\theta$. Discrete R-charge of pseudo-moduli
superfield $X$  is $2 R_d(\theta) \mod n$ and that of any other superfield is
just the subscript of that field.

%%%%%%%%%%
\section{Some trivial examples}\label{s:deformOR}
Adding a completely different and decoupled sector to any one of F-term SUSY
breaking models with U(1) R-symmetry (we will call it as old sector), we can
break U(1) R-symmetry to a $Z_n$ R-symmetry explicitly. Value of $n$ is
controlled by the new sector.  If we find a new sector (or in other words, a
$Z_n$ R-symmetry) for which no new term to the old sector is allowed even though
 R-symmetry becomes weaker, then SUSY breaking conditions coming from the old
sector will not alter. In this way, we get models of F-term SUSY breaking with a
discrete R-symmetry in generic theories.

To illustrate our idea, we consider the famous O'R model as an
example
of the old sector:
\begin{equation}\label{orafati}
 W= f X+a \phi_0 \phi_2+\frac{1}{2} b \phi_0^2 X.
\end{equation}
The above superpotential is generic with a U(1) R-symmetry where
$R(X)=R(\phi_2)=2,\,\,R(\phi_0)=0$ and a $Z_2$ internal symmetry under which
$X$ transforms trivially whereas $\phi$'s transform non-trivially. Let's now add
five new fields $\phi_0',\phi_1',\phi_2',\phi_3',\phi_4',$ to the old sector
and get the following superpotential.
\begin{eqnarray}\label{deform.orafati}
 W &=& f X+a \phi_0 \phi_2+\frac{1}{2} b \phi_0^2 X  + 
\frac{1}{2}\lambda'_{002}\phi_0^{'2} \phi_2'
+ \lambda'_{034}\phi_0' \phi_3' \phi_4'+ \frac{1}{2}\lambda'_{011}\phi_0'
\phi_1^{'2} \nonumber\\
&+& \frac{1}{6}\lambda'_{444}\phi_4^{'3}+ \lambda'_{124}\phi_1' \phi_2' \phi_4'+
\frac{1}{2}\lambda'_{133} \phi_1'\phi_3^{'2} + 
\frac{1}{2}\lambda'_{223}\phi_2^{'2} \phi_3'.
\end{eqnarray}
This superpotential does not have a U(1) R-symmetry, as can be easily checked. 
However,   spontaneous SUSY  breaking still occurs because the F-terms,
$F_X$ and  $F_{\phi_2}$, are not changed due to the inclusion of the new
terms. The above superpotential is  generic with the
following three symmetries.
\begin{enumerate}
 \item
A $Z_5$ R-symmetry with $R_d(\theta)=1$.
\item
A $Z_2$ internal symmetry under which all the $\phi_i$'s transform
non-trivially whereas 
$X$ and $\phi'_\alpha$'s transform trivially.
\item
A $Z_3$ internal symmetry under which all the $\phi'_\alpha$'s transform as 
$\phi'_\alpha\to \omega\phi'_\alpha$  whereas the remaining fields are
invariant.
\end{enumerate}

We can get different variations of the above model easily, for a
$Z_n$ R-symmetry with  $n\geq5$ and $R_d(\theta)=1$ as  follows.
\begin{eqnarray}\label{deform.OR.gen}
W &=& f X+a \phi_0 \phi_2+\frac{1}{2} b \phi_0^2 X  +
\frac{1}{6}\lambda'_{\alpha \beta \gamma}\phi_\alpha' \phi_\beta' \phi_\gamma'
\end{eqnarray}
where $\lambda'_{\alpha \beta \gamma}\neq0$ when $\alpha +\beta +\gamma=2 \mod
n$.
 Thus we have proved that there exists a large number of generic (trivial)
models where existence of a $Z_n$ R-symmetry is sufficient to break SUSY via
F-terms. One can also use this technique to other SUSY breaking models
\cite{Shih:2007av} with $U(1)$ R-symmetry.

 One side comment about these models. Old and new sectors  of any
one of these models can communicate to each other through gauge interactions if
we gauge some of the internal symmetries. For example, we can easily promote
the fields $\phi$ and $\phi'$  to transform under adjoint representation and
promote the field $X$ to remain invariant  under a gauge group, without
forbidding any term of the old and new sectors.  However, these models will then
 no longer be WZ models.

 Now, we can ask whether it is possible to construct generic F-term SUSY
breaking models  with the following charateristics: ($a$) there is no U(1)
R-symmetry  in the superpotential; ($b$) the superpotential cannot be subdivided
 into two disjoint sectors/parts.  In the rest of the paper, we show examples of
models  with all these characteristics.

%%%%%%%%%%
\section{A non-trivial example with some variations}\label{s:nontrivial}
We consider a renormalizable WZ model with $Z_{26}$ R-symmetry and
$R_d(\theta)=25$. We also consider that other than $X$, there are $\phi_3$,
$\phi_6$, $\phi_8$, $\phi_9$, $\phi_{11}$, $\phi_{12}$ and $\phi_{13}$ fields in
the theory. So we will have the following generic superpotential.
\begin{eqnarray}\label{e:example}
W_1 &=& f X + M_{11,13} \phi_{11} \phi_{13} + \frac{1}{2} M_{12,12}  \phi_{12}^2
+ \frac{1}{2} N_{13,13} X \phi_{13}^2+ \lambda_{3,8,13} \phi_3 \phi_8
\phi_{13}\nonumber\\
&&  + \lambda_{3,9,12} \phi_3 \phi_9
\phi_{12} + \frac{1}{2} \lambda_{6,6,12} \phi_6^2 \phi_{12} +
\frac{1}{2}\lambda_{6,9,9} \phi_6 \phi_9^2  + \frac{1}{6} \lambda_{8,8,8}
\phi_{8}^3
\end{eqnarray}
where, without loss of generality we can take  all the parameters, except
$\lambda_{8,8,8}$, to be real and positive. The above superpotential does not
have a U(1) R-symmetry. With
$\lambda_{3,9,12}=0$, there is a U(1) R-symmetry with the following R-charge
assignments.
\begin{eqnarray}
\begin{array}{|c|c|c|c|c|c|c|c|}
\hline
X & \phi_3 & \phi_6 & \phi_8 & \phi_9& \phi_{11} & \phi_{12}& \phi_{13}\\
\hline
\vphantom{\frac{\frac{l}{y}}{\frac{l}{y}}} 2 & \frac{4}{3} & \frac{1}{2}
& \frac{2}{3} & \frac{3}{4} & 2& 1&0\\
\hline
\end{array}
\end{eqnarray}
But then the superpotential is not generic. For $\lambda_{3,9,12}\ne0$, this
U(1) R-symmetry is explicitly broken.

There is spontaneous SUSY breaking. This can be easily realized by observing
the following F-terms,
\begin{eqnarray}\label{fx}
-F_X^* &=& f + \frac{1}{2}  N_{13,13} \phi_{13}^2 \\ \label{fphi11}
-F_{\phi_{11}}^* &=& M_{11,13} \phi_{13}.
\end{eqnarray}
Notice that vacuum expectation values (VEVs) of $F_X$ and $F_{\phi_{11}}$ terms 
cannot be simultaneously zero.

 We can vanish all other F-terms for any value of
$\phi_{13}^{(0)}$ (VEV of $\phi_{13}$) by choosing appropriate VEVs of other
fields. Now minimum of
scalar potential depends on $\phi_{13}^{(0)}$. Like the O'Raifeartaigh model
\cite{O'Raifeartaigh:1975pr} we have two cases, (a)~for
$y=\frac{f N_{13,13}}{M_{11,13}^2}<1 $, minimum is at $\phi_{13}^{(0)}=0$,
whereas (b)~for $y>1$, minimum is at  $\phi_{13}^{(0)}= \pm i\frac{M_{11,13}}
{N_{13,13}}\sqrt{2(y-1)}$. Hence we have a vacuum where supersymmetry as well as
discrete R-symmetry get spontaneously broken. 

Tree level scalar potentials of SUSY breaking often have flat directions
\cite{Ray:2006wk,Sun:2008nh}. For example, for the case of $y<1$, minimum of
tree level potential is independent of $X^{(0)}$ and $\phi_3^{(0)}$. So, it is
necessary to calculate 1-loop correction to check whether these flat directions
are lifted or not. One loop correction is given by Coleman-Weinberg
(CW) \cite{Weinberg:1973am} potential,
\begin{equation}\label{e:cw}
V_{CW} =\frac{1}{64 \pi^2}
\left(\mbox{tr}\left(M_B^4\log\frac{M_B^2}{\Lambda_{\rm cutoff}^2}\right)
-\mbox{tr}\left(M_F^4 \log\frac{M_F^2}{\Lambda_{\rm cutoff}^2}\right)\right),
\end{equation}
where $M_B$ and $M_F$ are mass matrices for scalar and fermion fields. Non-zero
eigenvalues $\lambda^F$ and  $\lambda^B$ of $M_F^2$ and $M_B^2$ respectively 
for $y<1$ are follows.
\begin{eqnarray}
\lambda_{1,\eta}^F&=& \lambda_{2,\eta}^F=\frac12 \Big(M_{12,12}^2+
2\lambda_{3,9,12} |\phi_3^{(0)}|^2 + \eta M_{12,12} \sqrt{ M_{12,12}^2 + 4
\lambda_{3,9,12}^2 |\phi_3^{(0)}|^2}\Big)\nonumber\\
\lambda_{3,\eta}^F&=& \lambda_{4,\eta}^F=\frac12 \Big(2M_{11,13}^2
+N_{13,13}^2 |X^{(0)}|^2 + 2 \lambda_{3,8,13}^2 |\phi_3^{(0)}|^2\nonumber\\
&& + \eta N_{13,13} |X^{(0)}|\sqrt{4 M_{11,13}^2 + N_{13,13}^2 |X^{(0)}|^2 + 4
\lambda_{3,8,13}^2 |\phi_3^{(0)}|^2}\Big)\nonumber\\
\lambda_{1,\eta}^B&=& \lambda_{2,\eta}^B=\lambda_{1,\eta}^F\nonumber\\
\lambda_{3,\eta_1,\eta_2}^B&=& \frac12\Big( \eta_2 fN_{13,13} + 2M_{11,13}^2
+N_{13,13}^2 |X^{(0)}|^2 + 2 \lambda_{3,8,13}^2 |\phi_3^{(0)}|^2
+\eta_1N_{13,13} \nonumber\\
&&  \sqrt{f^2 + |X^{(0)}|^2 (2 \eta_2 f N_{13,13}+ 4M_{11,13}^2
+N_{13,13}^2 |X^{(0)}|^2 + 4 \lambda_{3,8,13}^2 |\phi_3^{(0)}|^2) }\Big),
\nonumber
\end{eqnarray}
where $\eta$, $\eta_1$ and $\eta_2$ denote $\pm1$. Putting these
eigenvalues to \Eqn{e:cw} and expanding $V_{CW}$  about 
$X^{(0)}=\phi_3^{(0)}=0$, we find
\begin{eqnarray}
V_{CW} &=& \mbox{const.} + m_{X^{(0)}}^2 |X^{(0)}|^2 + m_{\phi_3^{(0)}}^2
|\phi_3^{(0)}|^2 + {\cal O}(|X^{(0)}|^4, |\phi_3^{(0)}|^4),
\end{eqnarray}
where 
\begin{eqnarray}
%V_0&=&  \nonumber\\
m_{X^{(0)}}^2&=& \frac{M_{11,13}^2N_{13,13}^2}{32 \pi^2}y^{-1}((1+y)^2\log(1+y)
-(1-y)^2\log(1-y) -2y)\nonumber\\
m_{\phi_3^{(0)}}^2&=& \frac{\lambda_{3,8,13}^2M_{11,13}^2}{64 \pi^2}
((1+y)\log(1+y)
+(1-y)\log(1-y)).
\end{eqnarray}
Constants $m_{X^{(0)}}^2$ and $m_{\phi_3^{(0)}}^2$ are positive and hence  after
addition of 1-loop correction, total scalar potential have a local minimum
at $X^{(0)}=\phi_i^{(0)}=0$.

We can get different variations of the above  model by adding more $\phi$
fields in the theory. For example, we can add any  number of fields from the
list $\{\phi_{16}, \phi_{19}, \phi_{21}, \phi_{22}, \phi_{25}\}$. In
this way we get 31 more models. If we add all the fields from the list, then
the superpotential takes the following form
\begin{eqnarray}
W &=& W_1 + M_{3,21} \phi_3 \phi_{21} + M_{8,16} \phi_8 \phi_{16}  +\frac{1}{2}
M_{25,25} \phi_{25}^2 + \lambda_{3,22,25} \phi_3 \phi_{22} \phi_{25} 
\nonumber\\
&& + \lambda_{6,19,25} \phi_6 \phi_{19} \phi_{25}+ \frac{1}{2} \lambda_{6,22,22}
\phi_6 \phi_{22}^2 + \frac{1}{2} \lambda_{8,21,21} \phi_8 \phi_{21}^2 +
\lambda_{9,16,25} \phi_9 \phi_{16} \phi_{25} \nonumber\\
&& + \lambda_{9,19,22} \phi_9 \phi_{19} \phi_{22} + \lambda_{12,13,25} \phi_{12}
\phi_{13} \phi_{25}+ \lambda_{12,16,22} \phi_{12} \phi_{16} \phi_{22} \nonumber
\\&&+ \frac{1}{2} \lambda_{12,19,19} \phi_{12} \phi_{19}^2 + \lambda_{13,16,25}
\phi_{13} \phi_{16} \phi_{25}.
\end{eqnarray}
Note that addition of these fields do not change $F_{X}$ and $F_{\phi_{13}}$ and
hence there is SUSY breaking.

%%%%%%%%%%
\section{Three series of models}\label{s:serises}
In this section we are going to show that there exists a large number of
non-trivial models where existence of a discrete R-symmetry is sufficient to
break supersymmetry. We consider the superpotentials are of the
following form with a $Z_{6k+2q}$ R-symmetry and $R_d(\theta)=6k+q$.
\begin{equation}\label{e:form}
W= X(f + \frac{1}{2} N_{3k+q,3k+q} \phi_{3k+q}^2) + M_{3k-q,3k+q} \phi_{3k-q}
\phi_{3k+q} + m(\phi) +\lambda(\phi),
\end{equation}
where $k,q$ are natural numbers, $\lambda(\phi)$ contains cubic terms which are
independent of  $\phi_{3k-q}$, and  $m(\phi)$ denotes quadratic terms for $\phi$
fields other than $\phi_{3k\pm q}$ fields. Note that the superpotentials of the
previous section are of the above form with $k=4$ and $q=1$. Due to this form 
of superpotentials, SUSY gets spontaneously broken. One can show this by
observing the following F-terms 
\begin{eqnarray}
-F_X^*&=& f + \frac12 N_{3k+q} \phi_{3k+q}^2\nonumber\\
-F_{\phi_{3k-q}}^*&=&M_{3k-q,3k+q}\phi_{3k+q}. 
\end{eqnarray}
Above equations are the same as \Eqs{fx}{fphi11} for $k=4$ and $q=1$.

%%%%%%%%%%%%%
\subsection*{Series I} 
In this series of models $k$ is multiple of four and $q=1$ i.e. superpotentials
 have a $Z_{6 k+2}$ R-symmetry with $R_d(\theta)=6 k+1$. Field content of this series
for any $k$ is given below.
\begin{equation}\label{e:seriesI}
\left\lbrace X, \phi_k, \phi_{2k-1}, \phi_{2k}, \phi_{2k+2}, \phi_{3k-1},
\phi_{3k}, \phi_{3k+1}, \phi_{2k \pm 4i}\Big(i=1,2,\ldots,\frac{k}{4}-1\Big)
\right\rbrace.
\end{equation}
Note that the model with $k=4$ have the same R-symmetry and $R_d(\theta)$
as the models given in the previous section. But this model is different from
those models  because it has different field content.

\begin{figure}
\begin{center}
\begin{picture}(200,50)(120,-15)
\Text(21,0)[l]{$\phi_{2k+2}$}
\Text(21,-8)[lt]{($-\frac{4}{3}$)}
\Line(48,0)(68,0)
\Line(58,0)(58,15)
\Text(58,16)[b]{$\phi_{2k+2}$}
\Text(69,0)[l]{$\phi_{2k-4}$}
\Text(69,-8)[lt]{($\frac{14}{3}$) {\bf A}}
\Line(98,0)(118,0)
\Line(108,0)(108,15)
\Text(108,16)[b]{$\phi_{2k}$}
\Text(119,0)[l]{$\phi_{2k+4}$}
\Text(119,-8)[lt]{($-\frac{10}{3}$)}
\Line(148,0)(168,0)
\Line(158,0)(158,15)
\Text(158,16)[b]{$\phi_{2k+4}$}
\Text(169,0)[l]{$\phi_{2k-8}$}
\Text(169,-8)[lt]{($\frac{26}{3}$) {\bf B}}
\Line(198,0)(218,0)
\Line(208,0)(208,15)
\Text(208,16)[b]{$\phi_{2k}$}
\Text(219,-8)[lt]{($-\frac{22}{3}$)}
\Text(219,0)[l]{$\phi_{2k+8}$}
\Line(248,0)(268,0)
\Line(258,0)(258,15)
\Text(258,16)[b]{$\phi_{2k+4}$}
\Text(269,0)[l]{$\phi_{2k-12}$}
\Text(269,-8)[lt]{($\frac{38}{3}$) {\bf C}}
\Line(298,0)(318,0)
\Line(308,0)(308,15)
\Text(308,16)[b]{$\phi_{2k}$}
\Text(319,-8)[lt]{($-\frac{34}{3}$)}
\Text(319,0)[l]{$\phi_{2k+12}$}
\Line(348,0)(368,0)
\Line(358,0)(358,15)
\Text(358,16)[b]{$\phi_{2k+4}$}
\Text(369,0)[l]{$\phi_{2k-16}$}
\Text(369,-8)[lt]{($\frac{50}{3}$) {\bf D}}
\Text(403,0)[l]{$\ldots$}
\end{picture}
\end{center}
\caption{Diagram representing some cubic terms in superpotentials for series I.
Values inside parentheses represent U(1) R-charges. For any field
$\phi_{2k+4i}$, U(1)  R-charge is $\frac{-12 i+2}{3}$ where $i$ is an
integer. The chain will be truncated at {\bf A, B, C,\ldots} for
$k=4,8,12,\ldots$}
\label{f:6kplus2} 
\end{figure}
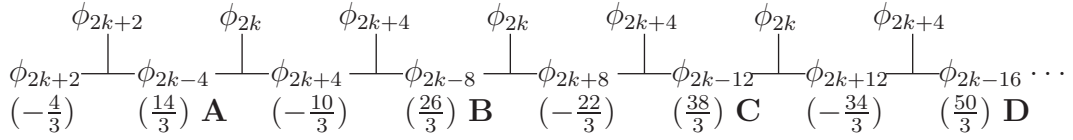

 To show that there is no U(1) R-symmetry in this series of models, we use
method of contradiction. If there were a U(1) R-symmetry in any model, existence
of terms $X \phi_{3k+1}^2$, $\phi_{3k}^2$, and $\phi_{2k}^3$ would imply
that $R(\phi_{3k+1})=0$, $R(\phi_{3k})=1$ and $R(\phi_{2k})=\frac{2}{3}$. From
the terms $\phi_k\phi_{2k} \phi_{3k}$, $\phi_{k} \phi_{2k-1} \phi_{3k+1}$  and
$\phi_{2k-1}^2 \phi_{2k+2}$, we could conclude $R(\phi_k)=\frac{1}{3}$, 
$R(\phi_{2k-1})=\frac{5}{3}$ and $R(\phi_{2k+2})=-\frac{4}{3}$.
Similarly we could construct a R-charge assignment chain for other fields as
shown in Fig.~\ref{f:6kplus2}. Now, for $k=4$ we have $2k-4=k$. But according
to the chain (at the point A) $R(\phi_{2k-4})=\frac{14}{3}\ne R(\phi_k)$. Hence
the superpotential for $k=4$ do not have a U(1) R-symmetry. Moving down the
chain, one can easily show that there is no U(1) R-symmetry in any model of this
series.

We are now going to prove that all the generic and renormalizable
superpotentials of this series of models are of the form as given
in \Eqn{e:form}.

There is no term quadratic in $X$ in the superpotentials because field content
for any $k$ does not contain the field $\phi_2$. Also, cubic term in $X$ is not
allowed by discrete R-symmetries.  Similarly, one can show that the terms
$\phi_{3k-1}^2\phi_2$ and $\phi_{3k-1}^3$ are also not allowed.

For a cubic term of the form $\phi_i\phi_j\phi_{3k-1}$ to exist, we need
$i+j=3k+1$. Without loss of generality, we can take $i\leq j$. From field
content given in \Eqn{e:seriesI}, we find $i\geq k$ and hence $j\leq 2k +1$.
Also one of them must be odd since their sum is odd. The only field having odd
discrete R-charge in this range is $\phi_{2k-1}$. But field content of any
model does not contain the field $\phi_{k+2}$ and so, $\lambda$-terms are
independent of the field $\phi_{3k-1}$. There is only one cubic term containing
$X$, $\frac{1}{2}  N_{3k+1,3k+1} X \phi_{3k+1}^2$, because discrete R-charges
of the $\phi$ fields lie between $k$ and $3k+1$. Thus superpotentials of this
series are of form  as given in \Eqn{e:form} and hence there is F-term SUSY
breaking.

Let's now give superpotential for $k=4$.
 \begin{eqnarray}
 W &=& f X +  M_{11,13} \phi_{11} \phi_{13}+  \frac{1}{2} M_{12,12} \phi_{12}^2
+ \frac{1}{2} N_{13,13} X \phi_{13}^2 + \lambda_{4,7,13} \phi_{4} \phi_{7}
\phi_{13} \nonumber\\
&&  +  \lambda_{4,8,12} \phi_{ 4} \phi_{ 8} \phi_{12} +  \frac{1}{2} \lambda_{4
,10,10} \phi_{ 4} \phi_{10}^2 +  \frac{1}{2} \lambda_{7,7,10}  \phi_{ 7}^2
\phi_{10} +  \frac{1}{6} \lambda_{8,8,8} \phi_{ 8}^3.
 \end{eqnarray}
One can explicitly verify that the above superpotential does not have a
U(1) R-symmetry yet there is F-term SUSY breaking.
%%%%%%% 
\subsection*{Series II }
Superpotentials of this series have $Z_{6 k+4}$ R-symmetry where $k$ is multiple
of 6 with starting value 12. Discrete R-charge of superspace co-ordinate
$\theta$ is $6k+2$ or $q=2$. Field content for any $k$ is given below.
\begin{eqnarray}
&&\Big\lbrace X, \phi_k,  \phi_{2k-6}, \phi_{2k-2}, \phi_{2k-1}, \phi_{2k},
\phi_{2k+1}, \phi_{2k+3}, \phi_{2k+6},\phi_{3k-2},\nonumber\\
&&  \phi_{3k}, \phi_{3k+2}, \phi_{2k \pm 6 i}
\Big(i=2,3,\ldots,\frac{k}{6}-1\Big). \Big\rbrace
\end{eqnarray}

To show that there is no U(1) R-symmetry, we have taken same the strategy as of
the earlier case. Let's first tabulate U(1) R-charges of first twelve fields
from the above list.
\begin{equation}
\begin{array}{|c|c|c|c|c|c|c|c|c|c|c|c|}
 \hline
X & \phi_k & \phi_{2k-6}& \phi_{2k-2} &\phi_{2k-1}&\phi_{2k} &
\phi_{2k+1}&\phi_{2k+3} & \phi_{2k+6} &
\phi_{3k-2} & \phi_{3k} & \phi_{3k+2}\\
\hline \vphantom{\frac{\frac12}{\frac12}}
2 & \frac{1}{3}&\frac{11}{3} & \frac{5}{3}& \frac{7}{6} & \frac{2}{3} &
\frac{1}{6}&
-\frac{5}{6}&-\frac{7}{3} & 2 & 1 & 0\\
\hline
\end{array}
\end{equation}
These fields are common to all models of this series. U(1) R-charges for the
fields $\phi_{3k+2}, \phi_{3k}, \phi_{3k-2},\phi_{2k-2}, \phi_{2k}$ and $\phi_k$
can be derived easily. We obtained R-charges for $\phi_{2k+1}, \phi_{2k-1},
\phi_{2k+3}, \phi_{2k-6}, \phi_{2k+6}$ from the terms
$\phi_{2k-2}\phi_{2k+1}^2$, $\phi_{2k+1}\phi_{2k}\phi_{2k-1}$,
$\phi_{2k-2}\phi_{2k-1}\phi_{2k+3}$, $\phi_{2k+3}^2\phi_{2k-6}$ and
$\phi_{2k-6}\phi_{2k}\phi_{2k+6}$ respectively.  U(1) R-charge assignment
chain for rest of the fields is  given in the Fig.~\ref{f:6kplus4}. From these
given information, one can easily show that there is no U(1) R-symmetry in any
model of this series.

\begin{figure}
\begin{center}
\begin{picture}(200,50)(180,-15)
\Text(119,0)[l]{$\phi_{2k+6}$}
\Text(119,-8)[lt]{($-\frac{7}{3}$)}
\Line(148,0)(168,0)
\Line(158,0)(158,15)
\Text(158,16)[b]{$\phi_{2k+6}$}
\Text(169,0)[l]{$\phi_{2k-12}$}
\Text(169,-8)[lt]{($\frac{20}{3}$) {\bf A}}
\Line(198,0)(218,0)
\Line(208,0)(208,15)
\Text(208,16)[b]{$\phi_{2k}$}
\Text(219,-8)[lt]{($-\frac{16}{3}$)}
\Text(219,0)[l]{$\phi_{2k+12}$}
\Line(248,0)(268,0)
\Line(258,0)(258,15)
\Text(258,16)[b]{$\phi_{2k+6}$}
\Text(269,0)[l]{$\phi_{2k-18}$}
\Text(269,-8)[lt]{($\frac{29}{3}$) {\bf B}}
\Line(298,0)(318,0)
\Line(308,0)(308,15)
\Text(308,16)[b]{$\phi_{2k}$}
\Text(319,-8)[lt]{($-\frac{25}{3}$)}
\Text(319,0)[l]{$\phi_{2k+18}$}
\Line(348,0)(368,0)
\Line(358,0)(358,15)
\Text(358,16)[b]{$\phi_{2k+6}$}
\Text(369,0)[l]{$\phi_{2k-24}$}
\Text(369,-8)[lt]{($\frac{38}{3}$) {\bf C}}
\Text(403,0)[l]{$\ldots$}
\end{picture}
\end{center}
\caption{Diagram representing some cubic terms in superpotentials for series
II. Values inside parentheses represent U(1) R-charges. For any field
$\phi_{2k+6 i}$, discrete R-charge is $\frac{-9 i+2}{3}$ where $i$ is an
integer. The chain will truncate at {\bf A, B, C, \ldots} for
$k=12,18,24,\ldots$. }
\label{f:6kplus4} 
\end{figure}

 There will be a $\lambda_{i,j,3k-2}$-term  only if $i+j=3k+2$. Without loss of
generality, we can take $i\le j$. Minimum value for $i$ is $k$. As there is no
field with R-charge $2k+2$, $i$ cannot be equals to $k$. Next higher value
of discrete R-charge is $k+6$ and hence $k+6\le i\le j < 2k-2$. In this range
discrete R-charges of the fields are multiple of 6. As $3k+2$ is not multiple of
6, there cannot a $\lambda$-term for $\phi_{3k-2}$. So, superpotentials of this
series also of the form as given in \Eqn{e:form} and  which in turn guarantee
spontaneous breakdown of SUSY.

Let's give first model of this series.
 \begin{eqnarray}
 W &=& f X +  M_{34,38} \phi_{34} \phi_{38}+  \frac{1}{2} M_{36,36}
\phi_{36}^2 +   \frac{1}{2}  X N_{38,38}\phi_{38}^2 + \lambda_{12,22,38}
\phi_{12} \phi_{22} \phi_{38} \nonumber
\\&&  +  \lambda_{12,24,36} \phi_{12} \phi_{24} \phi_{36} +  \frac{1}{2}
\lambda_{12,30,30} \phi_{12} \phi_{30}^2 +  \lambda_{22,23,27} \phi_{22}
\phi_{23} \phi_{27} +  \frac{1}{2} \lambda_{22,25,25}  \phi_{22} \phi_{25}^2 
\nonumber
\\&& +  \lambda_{23,24,25} \phi_{23} \phi_{24} \phi_{25} +  \frac{1}{6}
\lambda_{24,24,24} \phi_{24}^3+\lambda_{18,24,30} \phi_{18} \phi_{24} \phi_{30}
+ \frac{1}{2}  \lambda_{18,27,27}  \phi_{18} \phi_{27}^2\nonumber\\&& +  
\frac{1}{2} \lambda_{18,18,36} \phi_{18}^2 \phi_{36} 
 \end{eqnarray}
One can explicitly verify that there is no U(1) R-symmetry and SUSY is
spontaneously broken.
%%%%%%%%%%%%%
\subsection*{Series III}
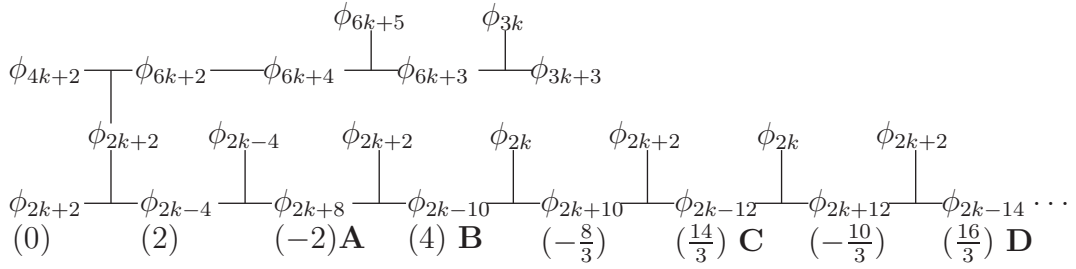
\begin{figure}
\begin{center}
\begin{picture}(200,85)(120,-15)
\Line(48,50)(66,50)
\Text(47,50)[r]{$\phi_{4k+2}$}
\Line(58,30)(58,50)
\Text(67,50)[l]{$\phi_{6k+2}$}
\Text(47,0)[r]{$\phi_{2k+2}$}
\Text(21,-8)[lt]{(0)}
\Line(48,0)(68,0)
\Line(58,0)(58,20)
\Text(63,20)[b]{$\phi_{2k+2}$}
\Text(69,0)[l]{$\phi_{2k-4}$}
\Text(69,-8)[lt]{(2) }
\Line(98,0)(118,0)
\Line(108,0)(108,20)
\Text(108,20)[b]{$\phi_{2k-4}$}
\Text(119,0)[l]{$\phi_{2k+8}$}
\Text(119,-8)[lt]{($-2$){\bf A}}
\Line(148,0)(168,0)
\Line(158,0)(158,20)
\Text(158,20)[b]{$\phi_{2k+2}$}
\Text(169,0)[l]{$\phi_{2k-10}$}
\Text(169,-8)[lt]{(4) {\bf B}}
\Line(198,0)(218,0)
\Line(208,0)(208,20)
\Text(208,20)[b]{$\phi_{2k}$}
\Text(219,-8)[lt]{($-\frac{8}{3}$)}
\Text(219,0)[l]{$\phi_{2k+10}$}
\Line(248,0)(268,0)
\Line(258,0)(258,20)
\Text(258,20)[b]{$\phi_{2k+2}$}
\Text(269,0)[l]{$\phi_{2k-12}$}
\Text(269,-8)[lt]{($\frac{14}{3}$) {\bf C}}
\Line(298,0)(318,0)
\Line(308,0)(308,20)
\Text(308,20)[b]{$\phi_{2k}$}
\Text(319,-8)[lt]{($-\frac{10}{3}$)}
\Text(319,0)[l]{$\phi_{2k+12}$}
\Line(348,0)(368,0)
\Line(358,0)(358,20)
\Text(358,20)[b]{$\phi_{2k+2}$}
\Text(369,0)[l]{$\phi_{2k-14}$}
\Text(369,-8)[lt]{($\frac{16}{3}$) {\bf D}}
\Text(403,0)[l]{$\ldots$}
%%%
%%%
\Line(95,50)(115,50)
\Text(115,50)[l]{$\phi_{6k+4}$}
\Line(145,50)(165,50)
\Text(165,50)[l]{$\phi_{6k+3}$}
\Line(155,50)(155,65)
\Text(155,65)[b]{$\phi_{6k+5}$}
\Line(195,50)(215,50)
\Text(215,50)[l]{$\phi_{3k+3}$}
\Line(205,50)(205,65)
\Text(205,65)[b]{$\phi_{3k}$}
\end{picture}
\end{center}
\caption{Diagram representing some cubic terms in superpotentials for $n=6k+6$.
Values inside parentheses represent U(1) R-charges.  For any field
$\phi_{2k+2i}$, U(1)  R-charge is $\frac{-2 i+2}{3}$ where $i$ is an integer.
Most of the fields from \Eqn{e:6kplus6} are also added to the chain so that one
can easily determine the U(1) R-charge easily.}
\label{f:6kplus6} 
\end{figure}

Superpotentials of this series have $Z_{6k+6}$ R-symmetry with
$R_d(\theta)=6k+3$ and $k= 8, 10, 12, 14, \ldots$. Field content for any
$k$ is given below.
\begin{eqnarray}
&&\Big\lbrace X, \phi_k, \phi_{2k-4},\phi_{2k}, \phi_{2k+2}, \phi_{2k+8},
\phi_{3k-3}, \phi_{3k}, \phi_{3k+3},\nonumber\\
&&    \phi_{4k+2}, \phi_{6k+2}, \phi_{6k+3}, \phi_{6k+4}, \phi_{6k+5},\phi_{2k
\pm 2 i} \Big(i=5,6,\ldots,\frac{k}{2}-1\Big) \Big\rbrace
\end{eqnarray}
If we demand that the superpotentials have an  U(1) R-symmetry, then we will
have a table (\Eqn{e:6kplus6}) and a chain (Fig.~(\ref{f:6kplus6})) of R-charge
assignments. From these inputs one can conclude that there is no U(1) R-symmetry
in any model of this series.
\begin{equation}\label{e:6kplus6}
\begin{array}{|c|c|c|c|c|c|c|c|c|c|c|c|}
 \hline
X& \phi_k& \phi_{2k}& \phi_{2k+2}& \phi_{3k-3}& \phi_{3k}&
\phi_{3k+3}&  \phi_{4k+2}& \phi_{6k+2}& \phi_{6k+3}& \phi_{6k+4}& \phi_{6k+5}\\
\hline \vphantom{\frac{\frac12}{\frac12}}
2 & \frac{1}{3} & \frac{2}{3}& 0 &2 & 1& 0 & \frac{2}{3} &\frac{4}{3} & 1&
\frac{2}{3}&\frac{1}{3}\\
\hline
\end{array}
\end{equation}

There will be a $\lambda$-term for $\phi_{3k-3}$ only if $i+j=3k+3 \mod (6k+6)$
where $i,j$ are discrete R-charges of fields coupled to it. Thus 
\begin{equation}
 i+j=3k+3 \, \mbox{ or }\, i+j=9k+9 
\end{equation}
For first case $k\le i,j\le 2k+3$. As there is no field with odd discrete
R-charge in this range, this possibility is ruled out. One can check that second
possibility is also ruled out. Hence superpotentials of this series are also of
the form given in \Eqn{e:form} and there are F-term SUSY breaking.

Let's give first model of this series, i.e. for $k=8$, so that one can verify
non-existence of a U(1) R-symmetry and spontaneous breakdown of SUSY.
  \begin{eqnarray}
 W &=& f X +  M_{21,27} \phi_{21} \phi_{  27} +  \frac{1}{2} M_{24,24}
 \phi_{24}^2 +  M_{50,52} \phi_{50} \phi_{52} +  \frac{1}{2} M_{51,51}
\phi_{51}^2\nonumber\\&&+ \frac{1}{2} X N_{27,27} \phi_{27}^2 +
\lambda_{8,16,24}\phi_{8} \phi_{16} \phi_{24} + \frac{1}{2} \lambda_{12, 12,24}
\phi_{12}^2 \phi_{24} +  \frac{1}{2} \lambda_{12,18,18} \phi_{12} \phi_{  18}^2 
\nonumber\\&&+  \frac{1}{6} \lambda_{16,16,16} \phi_{ 16}^3 +
\lambda_{16,34,52}\phi_{16} \phi_{34} \phi_{52} + \lambda_{18,34,50}
\phi_{18} \phi_{34} \phi_{50} + \lambda_{24,27,51}\phi_{24} \phi_{27} \phi_{51}
 \nonumber\\&& +\frac{1}{6} \lambda_{34,34,34} \phi_{34}^3 +  \frac{1}{2}
\lambda_{50,53,53} \phi_{50} \phi_{53}^2 +  \lambda_{51,52,53}\phi_{51} \phi_{
 52} \phi_{53} +\frac{1}{6} \lambda_{52,52,52} \phi_{52}^3.  
 \end{eqnarray}

In the above we have given only three series of models. However one can
construct many series of such models. 
%%%%%%%%%%
\section{Conclusions}
We have shown that there  exists a large number of generic and
renormalizable Wess-Zumino models where existence of a $Z_n$ R-symmetry is
sufficient to break SUSY spontaneously. And it is well known that if there is no
R-symmetry in WZ models with canonical K\"{a}hler potential,  then global minima
always preserve SUSY. So, existence of a $Z_n$ R-symmetry in a generic and
renormalizable Wess-Zumino model is a necessary condition for F-term SUSY
breaking. However, for even $n$ with $R_d(\theta)=\frac n2$, one cannot have
models of SUSY breaking because for these cases superpotentials as a whole
transform trivially and terms which are allowed or forbidden by these
R-symmetries can always be reproduced by some internal symmetries.

%%%%%%%%%%
\paragraph{Acknowledgments: }We thank Palash B Pal and Gautam Bhattacharyya for
discussions and valuable suggestions.

\end{document}